\documentclass[twocolumn,showpacs]{revtex4}

\usepackage{graphicx}

\begin{document}

\title{Comparison of unitary transforms}
\author{Erika Andersson$^{(a)}$, Igor Jex$^{(b)}$, and Stephen M. Barnett$^{(a)}$}
\date{\today}
\affiliation{(a) Department of Physics, University of Strathclyde, Glasgow G4 
0NG, UK\\
(b) Department of Physics, FNSPE, Czech Technical University,\\
Prague, B\v rehov\'a 7, 115 19 Praha, Czech Republic}
\begin{abstract}
We analyze the problem of comparing unitary transformations. The task is 
to decide, with minimal resources and maximal reliability, whether two given 
unitary transformations are identical or different. It is possible to make such 
comparisons without obtaining any information about the individual
transformations. Different comparison strategies are presented and 
compared with respect to their efficiency.
With an interferometric setup, it is possible to compare two unitary 
transforms using only one test particle. Another strategy makes use of
a two-particle singlet state. This strategy is more efficient than using a 
non-entangled two-particle test state, thus demonstrating the benefit of
entanglement. Generalisations to higher dimensional transforms 
and to more than two transformations are made.
\end{abstract}

\maketitle
\begin{widetext}
\section{Introduction}
Various tasks of quantum information processing require the use of quantum 
networks which realise desired quantum operations on a chosen input state.
The networks may be formed by arrays of elementary gates, 
realising single or two particle operations. 
With increasing complexity, it will become 
important to locate a possible error that can cause the malfunctioning of 
the whole network. The error, or erroneous gate, could be found by measuring
its properties using some estimation strategy \cite{pres,acin1,acin2,janz,
dari1,dari2,dari3,aha,cir}. 
The estimation of a whole set of gates, however, might be rather time consuming.
Moreover, we may not be interested in what the error is, only if there is an 
error or not. Therefore, it will be better to perform a simpler check only
to see whether the gate performs correctly or not. This leads us to consider 
testing whether two gates are the same or different by {\it comparing} them
to each other. Similarly 
it might be favorable to test gates on a production line when a master gate 
is available. The produced gates could then simply be compared to the master 
gate, and if a deviation in the behaviour of the gate is found, it will be 
discarded.  Transformation comparison is closely related to state 
comparison \cite{steve}.

In the following we will discuss strategies for comparing unitary 
transforms. The choice of comparison strategy will naturally
depend on what is already known about the transforms, and also on the 
resources available for the test. In the present work, a restriction is made to 
when each transform may be used only once, and to when the transforms are 
completely unknown. Other situations will be commented upon at the end.
The focus will also be on unambiguous transformation
comparison, meaning that whenever a result ``different'' or ``same'' is 
obtained, the result is also correct. In general, this forces us to accept 
inconclusive comparison outcomes, giving no information \cite{idp}. 
It is also possible to compare unitary transformations using other types
of strategies, such as minimum-error comparison strategies. We will
return to this possibility at the end.

It seems reasonable to assume that any unambiguous, minimum-error, or any 
other type of comparison strategy, would have to involve the action of the 
transforms to be compared upon some test state, followed by a measurement 
(or measurements) of the changed test state. To be more specific, we have to
build a network where each transform occurs once, feed a test quantum state 
through the network, and then measure the output state. The transforms may 
be used in any way inside the network; in parallel, acting on 
different qubits or particles, in series, acting on a particle one transform 
after the other, or possibly even in the form of controlled-U operations. 
This picture includes the possibility of entanglement with an environment,
for example, the case where two transforms U and V act in some way on two
particles, which may be entangled with other particles (or with each other).
To be general, one should also allow for sequential measurements, with actions 
conditional on previous measurement results. As we allow each transform to be 
used only once, sequential measurements will be of limited use.

To be able to say anything definite and unambiguous 
about whether the transforms are the same or different, we
have to choose the test state and the network in such a way that certain
output results are {\it unambiguously} associated with the transforms being
the same or different. From this point of view, it is clear that, if the
unitary transforms are completely unknown, a comparison can never give the 
definite answer that the transforms (two or more) are identical. To obtain
this information, we would have to reliably distinguish between the case
when the transforms are exactly the same, and when they are arbitrarily
close, but still not identical. That this is not possible can be seen by
a continuity argument similar to the one for state comparison \cite{steve}. 
The transforms being the same would have to be associated with the output 
state lying in a certain known subspace $H_{id}$ of the total allowed 
state space. When the transforms are almost, but not completely identical, 
the output state $|\Psi_{diff}\rangle$ will lie almost, but not completely, 
in this subspace:
\begin{equation}
|\Psi_{diff}\rangle = \alpha |\Psi_{id}\rangle + 
\beta |\Psi^{\perp}\rangle ,
\end{equation}
where $|\Psi_{id}\rangle $ lies in the subspace $H_{id}$,
and $|\Psi^{\perp}\rangle $ 
lies in the complementary subspace. By making the transformations arbitrarily
close, $\beta$ can be made arbitrarily small, and there is no measurement
that would distinguish unambiguously between $|\Psi_{id}\rangle$ and 
$|\Psi_{diff}\rangle$ for any arbitrarily small $\beta$. Such a measurement
would have to contain a measurement operator $\Pi_{id}$ for which, with
a finite success probability $p_{same}$ to detect the transforms as ``same'',
\begin{equation}
\langle \Psi_{id}|\Pi_{id}|\Psi_{id}\rangle =p_{same} \qquad{\rm and}\qquad
\langle \Psi_{diff}|\Pi_{id}|\Psi_{diff}\rangle =0 \quad\forall \beta .
\end{equation}
This forces the probability $p_{same}$ to be zero, meaning that we can 
never unambiguously
determine a set of transformations to be identical, unless we have
additional information about the transforms, for example, that they belong
to a certain given set of transforms.

In contrast, it is possible to obtain the unambiguous knowledge that the
transformations are different. To treat the problem in its full generality,
with the transforms to be compared appearing in an arbitrary network
as described above, is difficult. We will limit the discussion to the case 
when the transforms appear in parallel, as shown for two transforms 
$U$ and $V$ in Fig. (\ref{fig:UVnet}). An input state is sent 
through a preparation network, then acted upon by $U$ and $V$ as shown,
and finally measured. We have to be able to associate certain outcomes
unambiguously with the transforms being different.
\begin{figure}
\center{\includegraphics[width=7cm,height=!]{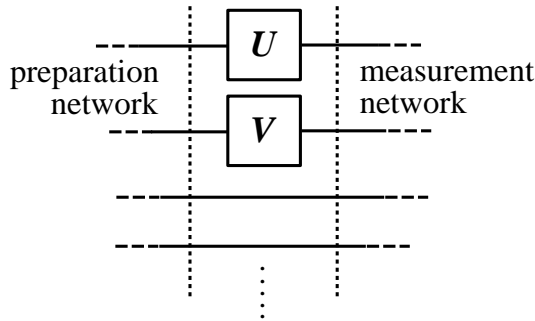}}
\caption{A general network for comparing two unitary transforms $U$ and $V$.}
\label{fig:UVnet}
\end{figure} 

We will first concentrate on comparing two single qubit 
transformations. These represent the simplest elements of quantum networks, 
and the ideas involved are demonstrated in a clear way. Here we present
different comparison strategies, one using only a single qubit as a test
state, and another strategy involving two qubits in a singlet state. These are
shown to succeed more often than a third strategy 
using two nonentangled qubits. The strategies will
then be generalised to comparison of more than two transforms, and to 
transforms of higher dimension. We end with conclusions.

\section{Comparing two unitary qubit transformations}
 
We will start with comparing two unknown single qubit transformations 
$U$ and $V$. A general way of parametrising $2\times 2$ unitary transforms
is to write them as $SU(2)$-transforms with a global phase,
\begin{equation}
U = e^{i\varphi_u} \left(\begin{array}{cr}
a_u & -b_u \\
b^*_u & a^*_u
\end{array}\right)~~~,~~~ 
V = e^{i\varphi_v} \left(\begin{array}{cr}
a_v &  -b_v \\
b^*_v & a^*_v
\end{array}\right)
\end{equation}
with $\vert a_{u(v)}\vert^2 + \vert b_{u(v)}\vert^2 = 1$. We see that four
real parameters suffice to determine each transform. The global phase will
matter whenever a unitary transform may or may not be applied to a physical 
system, such as in an interferometric setup, or in a controlled-U gate.
For this reason, we cannot regard transforms which have different global
phases, but otherwise are identical, as equivalent.
Surprisingly enough, it is found that a single particle test state is 
enough to test, in a single run, for differences in three of these four 
degrees of freedom. We will also consider the effect of entanglement using 
a two particle singlet test state, contrasting this method to a two particle
non-entangled strategy.

\subsection{Single qubit strategy}

Assume that, apart from two single copies of $U$ and $V$, only one
test qubit is available. Let us for the moment think about
photons, with the two qubit basis states being horizontal and vertical
polarisation, although the discussion applies to any two-level system.
A strong reason for considering what may be done with only one test
particle is that any unitary transform, acting on one photon, may be
achieved with linear optical elements. This means that it will be
possible to realise in an experiment any single-qubit
test strategy.

We will start by considering one intuitive example strategy, then generalising
the argument to obtain an optimal single particle comparison strategy.
As shown in Fig. (\ref{fig:onephot}), we may for example split the photon 
with a beam splitter, not necessarily 50/50, let $U$ and $V$ act in one 
path each, recombine the two paths with another beam splitter, not necessarily
identical to the first, and detect in which path and with what polarisation 
the photon exits. Phase shifts, not included in the figure, may be added 
at any point in the network.
\begin{figure}
\center{\includegraphics[width=6cm,height=!]{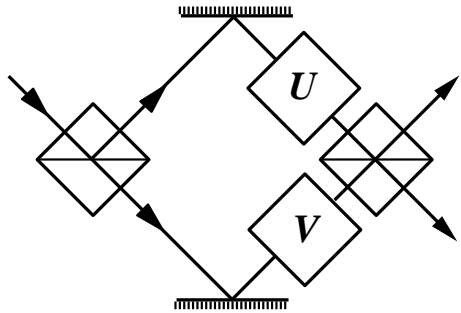}}
\caption{ A network for comparing $U$ and $V$ with a single test particle. 
$U$ and $V$ are acting in each arm of an interferometer.}
\label{fig:onephot}
\end{figure} 
The first beam splitter can be thought of merely as a state preparation
device. A general single photon test state is
\begin{equation}
\vert\psi_{in}\rangle = c {1\choose 0}_u + d {0\choose 1}_u + 
e {1\choose 0}_v + f {0\choose 1}_v,
\label{eq:in}
\end{equation}
where ${1\choose 0}_{u(v)}, {0\choose 1}_{u(v)}$ are the basis states of 
the photon in the paths going through $U$ and $V$, and $|c|^2+|d|^2+|e|^2+
|f|^2=1$. Any state of this form can be prepared by a linear state preparation
network built with polarising and non-polarising beam splitters, phase
shifts and polarisation rotations.
After $U$ and $V$ acting in the paths, the state is given by 
\begin{eqnarray}
\vert\psi_{out}\rangle &=& U\left[c {1\choose 0}_u + 
d {0\choose 1}_u \right] + 
V \left[ e {1\choose 0}_v + f {0\choose 1}_v\right]\nonumber\\
&=& e^{i\varphi_u} {{a_u c - b_u d}\choose{b^*_u c+ 
a^*_u d}}_u + e^{i\varphi_v} 
{{a_v e - b_v f}\choose{b^*_v e+ a^*_v f}}_v.
\label{eq:genstate}
\end{eqnarray}
To get a unique signal of dissimilarity we have to project on measurement 
states which are orthogonal to the output state obtained when $U=V$, i.e., 
to the state
\begin{equation}
\vert\psi_{out,U=V}\rangle = e^{i\varphi} 
{{a c - b d}\choose{b^* c+ a^* d}}_u 
+ e^{i\varphi} {{a e - b f}\choose{b^*e + a^* f}}_v .
\label{eq:samestate}
\end{equation}
The measurement is, in general, described by a probability operator measure
(POM) strategy, but, in this case, only a simple projective measurement using
a beam splitter (and phase shifts) will be required.
We can choose for instance $d=f=0$ and $c=e=1/\sqrt{2}$, so that
\begin{equation}
\vert\psi_{in}\rangle = \frac{1}{\sqrt{2}} \left[ {1\choose 0}_u + 
{1\choose 0}_v \right],
\label{eq:inex}
\end{equation}
representing a horizontally polarised photon, split equally between
the two paths. When $U=V$, the output state is 
\begin{equation}
\vert\psi_{out,U=V}\rangle = \frac{e^{i\varphi}}{\sqrt{2}} 
\left[ {a\choose b^*}_u + {a\choose b^*}_v \right] .
\end{equation}
This will always be orthogonal to the states
\begin{eqnarray}
\vert\psi_{1}\rangle &=& \frac{1}{\sqrt{2}} \left[ {1\choose 0}_u - 
{1\choose 0}_v \right]\nonumber\\
\vert\psi_{2}\rangle &=& \frac{1}{\sqrt{2}} \left[ {0\choose 1}_u - 
{0\choose 1}_v \right].
\label{eq:diffstates}
\end{eqnarray}
Whenever the photon is found exiting in either of these two states, $U$ and
$V$ must have been different.
In practice, the detection can be effected by recombining the two paths
by a second non-polarising beam splitter with the matrix
\begin{equation}
{1\over\sqrt{2}}\left(\begin{array}{cr}
1 & 1 \\
1 & -1
\end{array}\right)
\end{equation}
written in the ($u,v$) basis, meaning that if $U=V$, the output state is
\begin{equation}
\vert\psi_{out}\rangle = e^{i\varphi} {a\choose b^*}_u.
\end{equation} 
Whenever the photon is exiting in the $v$ path, $U$ and $V$ must have been
different. Note that even if $U$ and $V$ are not identical, the photon
may be found in path $u$. 
We can calculate the success rate for detecting the two transformations as 
different. The result reads
\begin{eqnarray}
P_{diff} &=&\langle\psi_{out}|\psi_1\rangle\langle\psi_1|\psi_{out}\rangle+
\langle\psi_{out}|\psi_2\rangle\langle\psi_2|\psi_{out}\rangle\nonumber\\
&=& \frac{1}{4} \left(| e^{i\varphi_u} a_u - 
e^{i\varphi_v}a_v\vert^2 + 
\vert e^{i\varphi_u} b_u - e^{i\varphi_v} b_v \vert^2 \right) .
\label{eq:pdiff}
\end{eqnarray}
If $U=V$, meaning $a_u=a_v$, $b_u=b_v$ and $\varphi_u=\varphi_v$, 
$P_{diff}$ is equal to zero as required for an unambiguous result.
If $U=V$, they will never be detected as different.
If we want to obtain an overall measure for the detection efficiency,
$P_{diff}$ should be averaged over all possible transforms $U$ and $V$:
\begin{equation}
\bar{P}_{diff}=\int dU\int dV P_{diff}\mu(U)\mu(V)={1\over 2}.
\label{eq:oneave}
\end{equation}
Here, $\mu$ has to be taken as the correct group-invariant 
measure, given for $SU(2)$ in \cite{group}. For $SU(2)$ with an
 additional overall phase $\varphi$, the correct measure is given by 
\begin{equation}
dU\mu(U)=d\theta d\alpha d\beta d\varphi {\cos 2\theta\over{8\pi^3}}, 
\end{equation}
where $a=\cos\theta e^{i\alpha}$, 
$b=\sin\theta e^{i\beta}$, $0<\theta<\pi/2$, $0<\alpha<2\pi$, 
$0<\beta<2\pi$, $0<\varphi<2\pi$, and similarly for $dV\mu(V)$.
This result, $\bar{P}_{diff} =1/2$, is due to the fact that the 
dimension of the subspace spanned by the states (\ref{eq:diffstates})
is two out of a total of four dimensions. Integrating over all possible
$U$ and $V$ corresponds to integrating over all possible output states
(\ref{eq:genstate}) in the four-dimensional Hilbert space. Projecting
onto two of these four dimensions obviously restricts the 
integration to these two dimensions, yielding $\bar{P}_{diff} =2/4=1/2$.

The result obtained is not dependent on the polarisation of the state 
$\vert\psi_{in}\rangle$ in Eq. (\ref{eq:inex}). Any other basis is equally 
suited. In general, what we have to find is measurement states
\begin{equation}
\vert\psi_{diff}\rangle = \tilde{c} {1\choose 0}_u +\tilde{d} {0\choose 1}_u 
+ \tilde{e} {1\choose 0}_v + \tilde{f} {0\choose 1}_v
\label{eq:meas}
\end{equation}
orthogonal to the state $|\psi_{out,U=V}\rangle$ in Eq. (\ref{eq:samestate}).
The fact that the orthogonality has to hold for all possible $a$, $b$ and 
$\varphi$ leads to the condition
\begin{equation}
\tilde{c}^*c+\tilde{e}^*e=\tilde{d}^*d+\tilde{f}^*f
=\tilde{c}^*d+\tilde{e}^*f=\tilde{d}^*c+\tilde{f}^*e=0,
\label{eq:ortcond}
\end{equation}
giving
\begin{equation}
ed=cf \quad {\rm and} \quad \tilde{e}\tilde{d}=\tilde{c}\tilde{f}.
\end{equation}
This means that the polarisation in the two paths has to be the same, but
any polarisation will do. The measurement states have to be adjusted 
according to (\ref{eq:ortcond}) to match the polarisation of the test state. 
The probability to detect a difference  will be optimal when the 
probabilities for the photon to be in each path are equal, both equal to 1/2.
Thus our intuitive first example of strategy with the test state 
(\ref{eq:inex}) proved to be one of the optimal one-photon test strategies
with a test state of the form (\ref{eq:in}).

Let us note that the single photon strategy fails to detect a difference
in $U$ and $V$ whenever $P_{diff}=0$ without the transforms being identical.
If we write $a=|a|e^{i\alpha}$ and $b=|b|e^{i\beta}$, with $|b|^2=1-|a|^2$,
the strategy fails to distinguish the transforms characterized by 
$(|a|,\alpha,\beta,\varphi )$ and $(|a|,\alpha +\delta ,
\beta+\delta,\varphi-\delta)$, for any $\delta$. Therefore, it effectively detects
differences in three parameters out of the four necessary to parametrize
a general single qubit unitary transformation. It is of course possible
to apply a mixed strategy, using two or more different polarisations for
the test state, for example, a choice of two different polarisations with
probability 1/2 each. One will then have a nonzero probability of detecting
any kind of difference in $U$ and $V$, even if the overall difference 
detection probability does not increase.

It is possible to further generalise the argument to any strategy where
at most one particle passes through $U$ or $V$, but allowing for
example entanglement with an environment. The general input state then reads
\begin{eqnarray}
|\psi_{in}\rangle &=&c|+\rangle_u|0\rangle_v|\phi_c\rangle_E+
d|-\rangle_u|0\rangle_v|\phi_d\rangle_E+
e|0\rangle_u|+\rangle_v|\phi_e\rangle_E\nonumber\\
&+&f|0\rangle_u|-\rangle_v|\phi_f\rangle_E+
g|0\rangle_u|0\rangle_v|\phi_g\rangle_E,
\label{eq:geninp}
\end{eqnarray} 
where $|+\rangle_{u,v}\equiv{1 \choose 0}_{u,v}$ and
$|-\rangle_{u,v}\equiv{0 \choose 1}_{u,v}$ refer to single photons in the
$u$ or $v$ path.
The vacuum state $|0\rangle_{u,v}$ denotes that no photons are present
in the $u$ or $v$ path. The states $|\phi\rangle_E$ are arbitrary but 
fixed environment states, describing the state of any particles which 
$U$ and $V$ will not explicitly act on, and may be selected
so as to optimise the strategy. The constants $c,d,e,f$ and $g$, whose 
absolute values sum to 1, are also chosen to optimise the strategy.
  
The transforms
$U$ and $V$ are then allowed to act in their respective paths, giving
an output state $|\psi_{out}\rangle$ exactly as before. By measuring
$|\psi_{out}\rangle$, we now have to judge whether $U$ and $V$ are
different or not. Any generalised measurement strategy may be 
described as a projective measurement in a higher-dimensional 
Hilbert space \cite{per}. A requirement for an unambiguous 
comparison strategy is therefore that there exists at least one 
measurement state
\begin{eqnarray}
|\psi_{diff}\rangle &=&\tilde{c}|+\rangle_u|0\rangle_v
|\tilde{\phi}_{\tilde{c}}\rangle_E+
\tilde{d}|-\rangle_u|0\rangle_v|\tilde{\phi}_{\tilde{d}}\rangle_E+
\tilde{e}|0\rangle_u|+\rangle_v|\tilde{\phi}_{\tilde{e}}\rangle_E\nonumber\\
&+&\tilde{f}|0\rangle_u|-\rangle_v|\tilde{\phi}_{\tilde{f}}\rangle_E+
\tilde{g}|0\rangle_u|0\rangle_v|\tilde{\phi}_{\tilde{g}}\rangle_E
\end{eqnarray} 
which is orthogonal to the output state $|\psi_{out,U=V}\rangle$ whenever
$U$ and $V$ are identical. Here, $|\tilde{\phi}\rangle_E$ denote
environment states, and $\tilde{c},\tilde{d},\tilde{e},\tilde{f}$ 
and $\tilde{g}$ are constants. There may exist several linearly
independent measurement states, as in the explicit example above.

As for the vacuum 
part of the general input state (\ref{eq:geninp}), the fact that the
orthogonality condition has to hold for all $\varphi_u=\varphi_v$ leads to
the requirement
\begin{equation}
\tilde{g}^*g\langle\tilde{\phi}_{\tilde{g}}|\phi_g\rangle = 0,
\label{eq:gcond}
\end{equation}
and although it is possible to have $g\neq 0$, after a quick consideration
one realises that $g=0$ is optimal: Due to condition 
(\ref{eq:gcond}), a nonzero $g$ will not contribute
to the success probability, but can only decrease it, since a nonzero 
$g$ forces the absolute values of $c$, $d$, $e$ and $f$ to be smaller.

To investigate whether entanglement with the environment may
be useful or not, let us not embark on any lengthy algebraic
calculation, but instead use an intuitive argument. In
the general input state (\ref{eq:geninp}), if 
$|\phi_c\rangle_E \neq |\phi_e\rangle_E$ or
$|\phi_d\rangle_E \neq |\phi_f\rangle_E$, then
it is possible to obtain which-path information, that is, 
knowledge about whether the single photon passed
through $U$ or $V$, by a measurement on the environment.
This inevitably degrades the success rate of the comparison,
which relies on our inability to determine whether the photon
passed through $U$ or $V$ in the same way which-path
information degrades the visibility in an 
interferometer. We are thus led to $|\phi_c\rangle_E =
 |\phi_e\rangle_E$ and $|\phi_d\rangle_E = |\phi_f\rangle_E$.
On the other hand, if $|\phi_c\rangle_E =
|\phi_e\rangle_E\neq|\phi_d\rangle_E =|\phi_f\rangle_E$,
this may be used to realise a mixed strategy employing 
either the test polarisation $|+\rangle$ or $|-\rangle$ with
desired probabilities. If $|\phi_c\rangle \perp |\phi_d\rangle$,
the selection is made by a detection
of the environment state in the $\{|\phi_c\rangle_E,
|\phi_d\rangle_E\}$ basis. This will not affect the comparison 
success probability, as it is independent of the polarisation 
of the test state. 

In applications such as dense coding \cite{dense},
entanglement may be used to enhance the distinguishability
of a set of unitary transforms. This is partly due to the enlarged
Hilbert space arising from having two qubits instead of one.
Here we have effectively enlarged the dimension of the avaliable
Hilbert space from two to four by introducing and extra degree of
freedom --- a choice of two paths --- to a single qubit  (photon),
without the explicit use of entanglement.
As another option, entanglement between two photons, one
passing through $U$, the other through $V$, may enhance the 
comparison success rate, not always, but in certain cases. This 
will be considered next.

\subsection{Two qubit singlet strategy}
\label{sec:singlet}

Let us consider a strategy which involves the use of a singlet state.
Experimentally, a singlet state of two photons may be prepared 
using parametric down-conversion. The test input state reads
\begin{equation}
\vert\psi_{in}\rangle =|\psi^-\rangle = \frac{1}{\sqrt 2} 
\left[ {1\choose 0 }_u 
{0\choose 1}_v - {0\choose 1}_u {1\choose 0}_v \right] .
\end{equation}
The singlet has the property that it remains invariant under the action 
of any two-particle transformation of the form $U_u\otimes U_v$.
\begin{figure}
\center{\includegraphics[width=7cm,height=!]{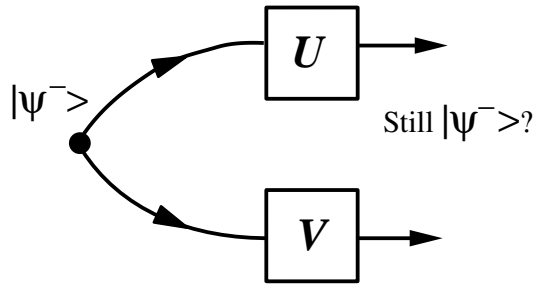}}
\caption{The transforms $U$ and $V$ are applied to a singlet test 
state. If the output state is not the singlet state, then $U$ and $V$ 
must have been different.}
\label{fig:singletnet}
\end{figure} 
The output state after passing through the transform $U_u\otimes V_v$
as in Fig. \ref{fig:singletnet} reads
\begin{eqnarray}
\vert\psi_{out}\rangle &=& e^{i(\varphi_u+\varphi_v )} [Re(a_u a^*_v + 
b_u b^*_v) \vert\psi_{in}\rangle +
i Im (a_u a^*_v + b_u b^*_v) \vert\psi^+\rangle + \nonumber\\
&&Re (b_u a_v - a_u b_v) \vert\Phi^+\rangle + 
i Im (b_u a_v - a_u b_v) \vert\Phi^-\rangle ] ,
\end{eqnarray}
where
\begin{equation}
\vert \Phi^{\pm}\rangle = \frac{1}{\sqrt 2} \left[ {1\choose 0 }_u 
{1\choose 0}_v \pm {0\choose 1}_u {0\choose 1}_v \right] 
\end{equation}
and
\begin{equation}
\vert\psi^{\pm}\rangle = \frac{1}{\sqrt 2} \left[ {1\choose 0 }_u 
{0\choose 1}_v \pm {0\choose 1}_u {1\choose 0}_v \right]
\end{equation}
are the Bell states. If $U=V$, then the output state is
\begin{equation}
|\psi_{out}\rangle=e^{2i\varphi}|\psi_{in}\rangle .
\end{equation}
If the output state is found in any Bell state other than the singlet, 
then this indicates that the transformations are different. This
requires a projection onto the antisymmetric versus the symmetric
subspace, and may be implemented with a non-polarising
beam splitter \cite{zeil}. 
The overall  success probability reads
\begin{equation}
P_{diff} = 1 - [Re(a_u a^*_v + b_u b^*_v)]^2.
\end{equation}
The averaged success rate, in analogy with Eq. (\ref{eq:oneave}), is
found to be 3/4. This is because the measurement
space, spanned by the three Bell states other than the singlet, is
now three-dimensional.
We should note that it is also possible
to start with any $|\psi_{in}\rangle$ in the subspace $\hat{1}-
|\psi^-\rangle\langle\psi^-|$ orthogonal to the singlet state, whereby a
detection of the output state in $|\psi^-\rangle$ indicates that the
transforms were different. In this case, however, the comparison 
success rate will be only 1/4, since the measurement space now 
has only one dimension out of four.

The singlet strategy can discriminate between all possible $SU(2)$ 
transformations, but cannot detect global phase differences 
$\varphi_u-\varphi_v$. This means that, like the single photon strategy,
it can detect differences in three of the four parameters. 
Two runs of the single photon
strategy with average success rate 1/2 each will have the same overall 
success rate, 3/4,  as one run of the singlet strategy. Thus it would seem
as if the use of entanglement carries limited advantage. However,
in the strictest
sense of the requirement ``only one use of $U$ and $V$ are allowed'', two
runs of the single photon strategy would not be permitted, even if the
average number of photons passing through $U$ and $V$ is one each.
Also, the singlet strategy is more efficient when it comes to
detecting small differences in the $SU(2)$ part of $U$ and $V$.
For a difference $\Delta\theta=\theta_u-\theta_v$, the single photon and
singlet strategies have difference detection probabilities of 
$\sin\Delta\theta$ and $\sin 2\Delta\theta$ respectively, meaning that 
for a small difference $\Delta\theta$, the singlet strategy is a factor of 
four better than 
the one-photon strategy, or a factor of two better than two single photon 
runs. On the other hand, the single-photon interferometric scheme is 
capable of detecting overall phase differences, where the singlet 
strategy fails. The choice of strategy depends on the type of 
difference to be detected. As already noted above, the single photon 
strategy introduces extra degrees of freedom to a single photon in 
allowing two different paths, so that the dimensionality of the test 
space for in the single photon strategy (four) equals that of  the singlet 
strategy (four).

\subsection{A non-entangled two particle strategy}
\label{sec:nonent}

We may also compare the singlet strategy to a non-entangled two particle
strategy based on state comparison \cite{steve}. Starting with two 
particles in the same state $|\Psi\rangle$, apply $U$ to one of them 
and $V$ to the other, as in Fig. \ref{fig:nonent}. 
\begin{figure}
\center{\includegraphics[width=7cm,height=!]{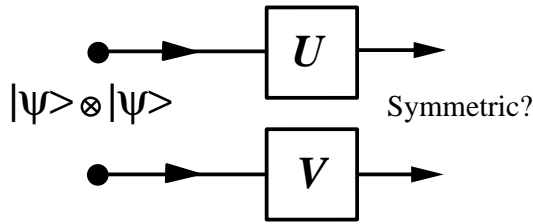}}
\caption{The transforms $U$ and $V$ are applied to two non-entangled 
test qubits. If their output state is found to be antisymmetric, then $U$ and 
$V$ must have been different.}
\label{fig:nonent}
\end{figure} 
There exists no unambiguous test telling us 
whether the particles are still the same state, but if they are found 
to be different, then we can infer that
$U$ and $V$ must have been different. The probability for this
strategy to succeed in detecting a difference in $U|\Psi\rangle$ and 
$V|\Psi\rangle$ is \cite{steve}
\begin{equation}
P_{diff}=\langle\Psi|\langle\Psi|\psi^-\rangle\langle\psi^-|\Psi\rangle|\Psi\rangle=
{1\over 2}(1-|\langle\Psi |U^\dagger V|\Psi\rangle|^2).
\label{eq:pdiffnonent}
\end{equation}
Choosing $|\Psi\rangle={1\choose 0}$ (the integrated overall success 
probability cannot depend on the state chosen if the transforms are 
completely unknown), and
using the facts that $Re(a_ua^*_v+b_ub^*_v)=Re(a_ua^*_v+b_u^*b_v)
\leq |a_ua^*_v+b_ub^*_v|=|\langle\Psi |U^\dagger V|\Psi\rangle|$ and 
$Re(a_ua^*_v+b_ub^*_v) \leq 1$,
this can be seen to be less than or equal to the success probability of
the singlet strategy. The difference in comparison with the singlet
strategy is a manifestation of gain due to entanglement. The overall
averaged success probability for one run of the non-entangled two-photon 
strategy is 1/4. This may be realised by integrating Eq. (\ref{eq:pdiffnonent})
directly, or by noting that detecting $U|\Psi\rangle$ and 
$V|\Psi\rangle$ as different corresponds to finding them in an 
antisymmetric state, in other words, in $|\psi^-\rangle$, which is one of four
orthogonal basis states.

An advantage of the single photon and non-entangled strategies is that
they are easily generalised to comparison of many transforms
and transforms of higher dimension, as will be seen below.

\section{Comparison of many unitary qubit transforms}

It is straightforward to generalise the single photon comparison strategy
to the case when $N$ $2\times 2$ unitary transformations are to be compared.
Denote the transforms by $U_i$, $i=1,...,N$, where
\begin{equation}
U_i = e^{i\varphi_i} \left(\begin{array}{cr}
a_i & -b_i \\
b^*_i & a^*_i
\end{array}\right).
\end{equation}
We want to find a strategy which tells us when all $U_i$ are not identical,
that is, when at least one $U_i$ differs from the rest.
In analogy with the comparison between two $2\times 2$ unitary 
transforms, use the single photon test state
\begin{equation}
\vert\psi_{in}\rangle = \frac{1}{\sqrt{N}} \sum_{i=1}^N{1\choose 0}_i,
\end{equation}
where ${1\choose 0}_i$ and ${0\choose 1}_i$ are the basis states (e.g.
polarisation) of a particle in the path going through $U_i$. This test
state may be obtained using a non-polarising beam splitter, which is
a linear device splitting an incident particle into $N$ different paths
with equal probability $1/N$ each, independent of the polarisation
\cite{igor}.
As when comparing two transforms,
the same freedom in choosing the polarisation applies here; as long as
the polarisation is the same in all paths, any polarisation may be used.
After action with $U_i$ in the different paths the output state reads
\begin{equation}
|\psi_{out}\rangle =\frac{1}{\sqrt{N}}\sum_{i=1}^N e^{i\varphi_i}
{a_i\choose b^*_i}_i.
\end{equation}
If all $U_i$ are identical, meaning $a_i=a, b_i=b$, and 
$\varphi_i=\varphi$ for all $i$, then
\begin{equation}
|\psi_{out,id}\rangle = \frac{e^{i\varphi}}{\sqrt{N}} 
\sum_{i=1}^N {a\choose b^*}_i = e^{i\varphi} 
\left[a \frac{1}{\sqrt{N}}\sum_{i=1}^N {1\choose 0}_i+ 
b^*\frac{1}{\sqrt{N}}\sum_{i=1}^N {0\choose 1}_i\right]
\equiv e^{i\varphi}(a|\Psi_{id}^1\rangle+b^*|\Psi_{id}^2\rangle).
\label{eq:nsame}
\end{equation}
This state is orthogonal to states of the form
\begin{equation}
|\Psi^{1-}_{ij}\rangle=\frac{1}{\sqrt{2}}\left[{1\choose 0}_i-
{1\choose 0}_j\right],
|\Psi^{2-}_{ij}\rangle=\frac{1}{\sqrt{2}}\left[{0\choose 1}_i-
{0\choose 1}_j\right], i \neq j.
\label{eq:ndiff}
\end{equation}
These states form an overcomplete basis for the subspace
orthogonal to the one spanned by states of the form (\ref{eq:nsame}).
Whenever the photon is detected in this orthogonal subspace, this
shows unambiguously that at least one of the $U_i$ was not identical to
the others. 
The detection can be effected by combining the $N$ paths with a second
non-polarising $2N$-port beam splitter, similar to the one used for the
test state preparation. The state $|\psi_{out,id}\rangle$ will then be mapped
onto
$e^{i\varphi}{a\choose b^*}_1,$
and whenever the photon is exiting in a path $i\neq 1$, we know that the
$U_i$ were not all the same.

The success probability to detect a difference reads
\begin{eqnarray}
P_{diff,N}&=&\frac{2}{N}\sum_{i,j=1;i>j}^N \langle\psi_{out}|
\Psi^{1-}_{ij}\rangle\langle\Psi^{1-}_{ij}|\psi_{out}\rangle+
\langle\psi_{out}|
\Psi^{2-}_{ij}\rangle\langle\Psi^{2-}_{ij}|\psi_{out}\rangle\nonumber\\
&=&\frac{1}{N^2}\sum_{i,j=1;i > j}^N
(|e^{i\varphi_i}a_i-e^{i\varphi_j}a_j|^2+|
e^{i\varphi_i}b_i-e^{i\varphi_j}b_j|^2).
\end{eqnarray}
To obtain the overall success probability to detect differences in $N$
transforms, we should integrate $P_{diff,N}$ over all possible transforms
$U_i$. 
Alternatively, we may use 
\begin{equation}
P_{diff,N}=1-\langle\psi_{out}|\Psi_{id}^1\rangle\langle\Psi_{id}^1|\psi_{out}\rangle -
\langle\psi_{out}|\Psi_{id}^2\rangle\langle\Psi_{id}^2|\psi_{out}\rangle.
\end{equation}
Integrating this over $U_i$, we obtain $\bar{P}_{diff,N}=1-1/N$.
Again, this is consistent with the fact that the dimension of the 
subspace spanned by the states in Eq. (\ref{eq:ndiff}) is $2N-2$, the 
dimension of the total Hilbert space being $2N$, so that  
$\bar{P}_{diff,N}=(2N-2)/2N=1-1/N$.
As the number of transforms grows, it is, loosely speaking, more likely 
that at least one of them differs from the rest, and our probability to 
detect a difference approaches unity. Of course, for any number of transforms
selected at random, the chance that they are all exactly equal is zero;
in this sense, the probability for randomly selected transforms to be 
different is one. But when the number of transforms increase, our 
probability of unambiguously detecting a difference also approaches unity.
  
In analogy with the non-entangled two particle comparison strategy
for two transforms in Sec. \ref{sec:nonent}, it is also possible to compare 
many unitary transforms 
by applying each transform 
to a copy of a state $|\Psi\rangle$, and then comparing the states.
Whenever all the transforms are identical, all the states will also
be identical, and the total state will be symmetric.
Thus, whenever the joint state of all the test particles
is found outside the totally symmetric subspace, the transforms
cannot all have been identical. (The particles may be distinguished
through their spatial locations, otherwise we could not apply one
unitary transform to each specific particle.)
The overall comparison success
rate $\bar{P}_{diff}$ will be equal to $1-D_s/2^N$, where $2$ is the
dimension of each unitary transform, $N$ is the number of transforms
to be compared and $D_s$ is the dimension of the totally
symmetric subspace for $N$ $2$-dimensional systems.

For example, for comparing three $2\times 2$ transforms, one may construct
the three-particle states
\begin{equation}
|\psi^0_{ij}\rangle=
{1\over \sqrt{2}}(|01\rangle_{ij}-|10\rangle_{ij})|0\rangle_k
\quad {\rm and} \quad
|\psi^1_{ij}\rangle=
{1\over \sqrt{2}}(|01\rangle_{ij}-|10\rangle_{ij})|1\rangle_k,
\end{equation}
where $i,j,k$ may be any permutation of 1,2,3, referring to the three 
particles, and $|0\rangle$, $|1\rangle$
are the single particle basis states. The projector onto the non-totally
symmetric subspace may then be formed as
\begin{equation}
\hat{1}-\hat{P}_{symm}={1\over 3}\sum_{i\neq j}\left(|\psi^0_{ij}\rangle
\langle\psi^0_{ij}|+|\psi^1_{ij}\rangle
\langle\psi^1_{ij}|\right).
\end{equation}
It follows that this method is related to pairwise comparison of the
particle states.

\section{Comparison of unitary transforms of higher dimension}

It is also possible to consider comparison of unitary transforms with a
dimension higher than $2\times 2$. Below we will briefly suggest strategies
for this. Physically, a $4\times 4$ transform
may describe a two-qubit operation, or a single-particle operation on
a single physical system with four dimensions. A ``mathematical''
comparison strategy derived for comparing $4\times 4$ unitary transforms
would work in both cases, but the physical implementations are different.
 
As before, we have to act with the 
transforms on a test state, which may be a multi-particle state,
selecting the test state and the way the transforms are used so that
certain outcomes indicate unambiguously that the transforms were different.
It is straightforward to generalise the single-particle strategy used above; 
we will sketch the extension here. Let us consider the case of two $M\times M$
unitary matrices $U$ and $V$, with matrix elements $a_{ij}$ and $b_{ij}$.
Denote the $M$ basis states in the $u$ and $v$ paths by $|i\rangle_{u,v}$,
where $i=1,...M.$ Using the test state
\begin{equation}
|\psi_{in}\rangle =\frac{1}{\sqrt{2}}(|1\rangle_u+|1\rangle_v),
\end{equation}
we may compare the first columns of $U$ and $V$. The test state
\begin{equation}
|\psi_{in}\rangle =\frac{1}{\sqrt{2}}(|i\rangle_u+|i\rangle_v)
\label{eq:multitest}
\end{equation}
compares the $i$th columns of $U$ and $V$. If a general comparison of
all columns is desired, then we may employ a mixed strategy, e.g. 
selecting the $M$ states (\ref{eq:multitest}) with probability $1/M$ each. 
The success probability to detect a difference in $U$ and $V$ for this
mixed strategy is
\begin{equation}
P_{diff}=\frac{1}{4M}\sum_{i,j=1}^M|a_{ij}-b_{ij}|^2.
\end{equation} 
If $U$ and $V$ are single particle transforms, 
then this is still a true single particle comparison strategy, 
meaning that it is possible to implement for example with
linear optics.
Otherwise, if $U$ and $V$ are multi-particle operations,
the test state is a multi-particle state, the basis states 
$|i\rangle$ being multi-particle states. 
It is a straightforward mathematical task to generalise this type of
comparison strategy to the case of more than two $M\times M$ transforms.
The averaged comparison success probability will be $1-1/N$, where
$N$ is the number of transforms.

The singlet strategy in Sec. \ref{sec:singlet} employed the concept of
an invariant subspace, as the singlet is invariant under $U\otimes U$.
When comparing $N$ $N$-dimensional unitary transforms, it is
possible to generalise this idea. A totally antisymmetric state of $N$
$N$-dimensional particles (a Slater determinant) is invariant
if the same unitary transform is applied to all $N$
particles. Detection of the system in any other state than the totally
antisymmetric state after application of one transform to each
particle therefore means that the transforms cannot all have been
identical. It is here understood that the Slater determinant refers to
the internal degree of freedom of the $N$ particles, which may be
distinguished because of different spatial location. Otherwise we
could not apply one specific transform to each particle. The dimension of
the total internal Hilbert space is therefore $N^N$, and, since there
is only one totally antisymmetric state, the averaged comparison
success probability is 1-1$/N^N$. This may be compared with the
averaged success probability of the single particle strategy, $1-1/N$. 
A non-entangled strategy based on state comparison succeeds
with a probability 1-$D_s/N^N$, where $D_s$ is the dimension of
the totally symmetric subspace. 

If the number of transforms to be compared is not equal to their
dimensionality, then we still 
have to look for the lowest dimensional 
invariant subspaces. 
We may choose a test state lying in the lowest
dimensional invariant subspace available, apply the 
transforms to be compared, and then check whether the output state 
remains in the same subspace or not. 
Strictly speaking the subspace does not even have to be invariant, it
will suffice if we are able to unambiguously predict in which subspace 
$H_{id}$ the output state will lie if the transforms are identical. An 
output state in the space complementary to $H_{id}$ will then indicate 
a difference in the transforms.
The comparison of two two-qubit transformations, for instance, would 
require us to work with the following four-particle states:
\begin{eqnarray}
\label{eq:nonsym}
\vert\psi_1 \rangle &=&\frac{1}{\sqrt 2} (\vert 1\rangle_1 \vert 0\rangle_2 
\vert 0\rangle_3 \vert 1\rangle_4 -  \vert 0\rangle_1 \vert 1\rangle_2 
\vert 1\rangle_3 \vert 0\rangle_4 )\nonumber\\
\vert\psi_2 \rangle &=&\frac{1}{\sqrt 2} (\vert 1\rangle_1 \vert 1\rangle_2 
\vert 0\rangle_3 \vert 0\rangle_4 -  \vert 0\rangle_1 \vert 0\rangle_2 
\vert 1\rangle_3 \vert 1\rangle_4 )\nonumber\\
\vert\psi_3 \rangle &=& \frac{1}{\sqrt 2} \vert 1\rangle_1 \vert 1\rangle_3 
(\vert 0\rangle_2 \vert 1\rangle_4 -  \vert 1\rangle_2 \vert 0\rangle_4 )\\
\vert\psi_4 \rangle &=& \frac{1}{\sqrt 2} \vert 0\rangle_1 \vert 0\rangle_3 
(\vert 0\rangle_2 \vert 1\rangle_4 -  \vert 1\rangle_2 \vert 0\rangle_4 )
\nonumber\\
\vert\psi_5 \rangle &=& \frac{1}{\sqrt 2} \vert 0\rangle_2 \vert 0\rangle_4 
(\vert 0\rangle_1 \vert 1\rangle_3 -  \vert 1\rangle_1 \vert 0\rangle_3 )
\nonumber\\
\vert\psi_6 \rangle &=& \frac{1}{\sqrt 2} \vert 1\rangle_2 \vert 1\rangle_4 
(\vert 0\rangle_1 \vert 1\rangle_3 -  \vert 1\rangle_1 \vert 0\rangle_3 ).
\nonumber
\end{eqnarray}
These states span the non-totally symmetric subspace ($|0\rangle$
and $|1\rangle$ are the qubit basis states). Its complementary subspace
is the symmetric subspace, which is ten-dimensional. A test state
lying either in the subspace spanned by the states (\ref{eq:nonsym}) or
in the totally symmetric subspace will remain in the same subspace
whenever the two transforms are identical. A detection of the output
state in the complementary subspace therefore unambiguously implies
that the transforms were different. Starting in the lower dimensional
non-totally symmetric subspace, however, will yield a higher comparison 
success probability,  $D_s/4^2=10/16$, as opposed to $1-10/16$ when 
starting in the ten-dimensional symmetric subspace.

The non-entangled strategy based on state comparison makes
use of the totally symmetric subspace as a starting space.
We may use it for comparing any number of unitary transforms of 
any dimension, by acting
with the transforms to be compared upon copies of a test state,
and then comparing all the test states. Although an entangled 
strategy may perform better, because it can use an invariant 
subspace with a lower dimension,
a non-entangled strategy may be easier to implement experimentally.
The single particle strategy is clearly the easiest to realise. Its success
probability also scales favorably as $1-1/N$
with the number of transforms.

\section{Some remarks on comparison of partially known transforms}

If we have additional information about the transforms, then we may use
this knowledge in selecting the test state and the way in which to use
the (single) copies of the transforms available, and in how to detect
differences and similarities from the output state. This is possible 
whatever strategy chosen, using a single particle or a many particle test 
state, for unambiguous and minimum-error strategies, or for any other
type of comparison strategy. The same freedom to select for example 
entangled test states applies.

It will always be possible
to use the universal comparison strategies considered above, but it may
be possible to improve on the success probability $P_{diff}$, and also
to obtain an unambiguous answer that the transformations are identical.
This is the case if the two transforms are known to be members of a given
set, e.g. $\{U_1,U_2\}$. There are then four possibilities for the output 
state, depending on whether $U=V=U_1$, $U=V=U_2$, $U_1=U\neq V=U_2$, or
$U_2=U\neq V=U_1$, and we have to distinguish the two former ones from
the two latter ones. In other words, we have to distinguish the
subspace spanned by the two output states when $U=V$ from the subspace
spanned by the output states when $U\neq V$. In general, these two
subspaces are nonorthogonal, and cannot be perfectly discriminated. But
in analogy with state discrimination, it will be possible to unambiguously 
distinguish these two possibilities, allowing for inconclusive results.
It will be possible to obtain an unambiguous answer not only that $U$ and 
$V$ are different, but also that they are identical. We could also choose 
to make a minimum-error measurement. 

If we are given the information that $U$ and $V$ are both controlled-NOT
gates, but not in which basis, then, if they have the same basis, $UV$
will be the identity operator. This implies that if $UV$ applied to a state
changes the state, the bases of the two controlled-NOT gates $U$ and $V$
must have been different, and this can be used for comparison.
This same argument may be useful when any cyclic transformations are to be
compared.

\section{Conclusions}

In this paper we have considered the situation where one is interested
in whether two or more unitary transforms are the same or different. 
No knowledge about the individual transforms is desired, 
only whether there is a difference or not. It is indeed
possible to {\it compare} the transforms without first estimating them. 
A ``trial'' copy of a transform may also be compared to a master copy. This 
can be used to test for defects in the ``trial'' copy, and may, in this
sense, be used as a way of transformation estimation. We have considered
unambiguous comparison strategies, which allows us to make confident
statements about the transforms.
When the transforms are completely unknown, it is not possible to
obtain an unambiguous result that the transforms are identical, but it is
possible to detect them as different. 

We have treated the case of comparing two $2\times 2$ unitary transforms
in depth, and made some generalisations.
Three types of unambiguous comparison strategies 
were derived: a single particle strategy, a singlet strategy
employing the concept of invariant subspaces, and a non-entangled
two particle strategy. Success probabilities are
calculated for these strategies.
Obvious extensions to the present work include a deeper treatment of
transformation comparison when there is knowledge about the transforms.

The conceptual connection to state comparison is obvious \cite{steve}.
One may also draw an analogy to the Deutsch-Jozsa algorithm for
determining whether a function is balanced or constant with a single 
application of the function \cite{deutsch}. Here, the function values are
not interesting, only whether the function is balanced or equal. 
Similarly, we have devised strategies for determining whether unitary 
transforms are different or not, without obtaining information about what the
transforms actually are. The feature that collective information may be 
obtained without knowledge of the parts of a system is characteristic
of many quantum algorithms vs. their classical counterparts. It is connected
to the possibility of choosing the measurement basis for a quantum system
more freely than for a classical system.

\begin{acknowledgments}
This work was supported by the UK Engineering and Physical Sciences
Research Council, the Royal Society of Edinburgh and the Scottish Executive
Education and Lifelong Learning Department. The financial support by GACR
202/01/0318 and EU IST-1999-13021 for I.J., and by the EU 
Marie Curie program, project number HPMF-CT-2000-00933 for E. A.
is gratefully acknowledged.
\end{acknowledgments}

\end{widetext}
\end{document}